\documentclass[aps,pre,preprint,superscriptaddress,showpacs]{revtex4-1}

\usepackage{amsmath, amsfonts, amsthm, amssymb, mathrsfs, enumerate, graphicx}

\begin{document}


\title{Helicalised fractals}


\author{Vee-Liem Saw}
\email[]{VeeLiem@maths.otago.ac.nz}
\affiliation{Division of Physics and Applied Physics, School of Physical and
Mathematical Sciences, Nanyang Technological University, 21
Nanyang Link, Singapore 637371}

\author{Lock Yue Chew}
\email[]{lockyue@ntu.edu.sg}
\affiliation{Division of Physics and Applied Physics, School of Physical and
Mathematical Sciences, Nanyang Technological University, 21
Nanyang Link, Singapore 637371}
\affiliation{Complexity Institute, Nanyang Technological University, Singapore}


\date{\today}

\begin{abstract}
We formulate the helicaliser, which replaces a given smooth curve by another curve that winds around it. In our analysis, we relate this formulation to the geometrical properties of the self-similar circular fractal (the discrete version of the curved helical fractal). Iterative applications of the helicaliser to a given curve yields a set of helicalisations, with the infinitely helicalised object being a fractal. We derive the Hausdorff dimension for the infinitely helicalised straight line and circle, showing that it takes the form of the self-similar dimension for a self-similar fractal, with lower bound of 1. Upper bounds to the Hausdorff dimension as functions of $\omega$ have been determined for the linear helical fractal, curved helical fractal and circular fractal, based on the no-self-intersection constraint. For large number of windings $\omega\rightarrow\infty$, the upper bounds all have the limit of 2. This would suggest that carrying out a topological analysis on the structure of chromosomes by modelling it as a two-dimensional surface may be beneficial towards further understanding on the dynamics of DNA packaging.
\end{abstract}


\maketitle


\section{Introduction}

Our study in the area of fractal geometry is originally motivated by the following goal: Given a smooth curve, replace it by another smooth curve that winds around it. After this is done, the resulting curve is replaced by yet another one that winds around it. What will the ultimate curve be after infinitely many iterations of this process?

For example if a straight line is replaced by a curve that winds around it, the result is a helix. The helix can then be replaced by a curve that winds around it, and so on. We shall call the process of replacing a given smooth curve by one that winds around it as \emph{helicalising the curve}. The curve that is produced after an infinite number of helicalisations would be a fractal \cite{fle1,fle2,tolsua1}. Notice however that this iterative helicalisation process does not result in a self-similar fractal as the $n$-th level is not some number of exact scaled copies of the $(n-1)$-th level. The Hausdorff dimension according to Ref. \cite{tolsua1} (with the general definition of the Hausdorff dimension found in Ref. \cite{text1}) is stated as
\begin{eqnarray}\label{Hausdorff}
D=-\frac{\ln{\omega}}{\ln{R}},
\end{eqnarray}
where $\omega>1$ is an integer representing the number of windings and $0<R<1$ is the ratio of the radius of windings of the $n$-th level to that of the $(n-1)$-th level, though a different expression for the Hausdorff dimension was provided by Refs. \cite{fle1,fle2} involving the pitch angle. This difference in their Hausdorff dimensions is due to some variations in the manner in which they generated their respective fractals, to specifically suit their own purposes. For instance, the motivation of Refs. \cite{fle1,fle2} came from studying wave propagation in such structures to understand the behaviour of elastic rods, whilst that in Ref. \cite{tolsua1} was to allow for his development of an algebraic structure of the fractal dimension.

We present our own definition on helicalising a curve in Section 3, formulated by extending the geometrical features of the self-similar discrete version of the fractal to a continuous form (the circular fractal is the discrete version of the curved helical fractal) which we discuss in Section 2. While Refs. \cite{fle1,fle2,tolsua1} dealt only with the fractal resulting from a straight helix, we are also interested in the fractal resulting from a curved one, by applying the iterative helicalisation procedure on a circle, since this provides comparisons between different helicalised fractals resulting from different starting curves. Our exact formulation followed by the subsequent analysis of the helicaliser based on the geometrical features of the discrete version differs from Refs. \cite{fle1,fle2,tolsua1} and may be employed as a standard reference for subsequent work in this area. In Section 4, we explicitly derive the Hausdorff dimension by relating the number of windings and the length of one winding as a cover. In addition, we also illustrate that by considering the scale factor in the limit of the infinite levels, the usual self-similar dimension expression is recovered, despite the fact that such helicalised fractals are not exactly self-similar. This formulation thus enables a relatively simple and tractable analysis to be performed on the fractals, by drawing directly from the intuitive geometrical meaning gleaned from the fractal construction.

Refs. \cite{fle1,fle2} stated that their Hausdorff dimension lies between 1 and 2, excluding the boundary values. This range is due to their geometrical considerations forbidding the overlap of the fractal curve. Ref. \cite{tolsua1} on the other hand did not provide any upper bound, though it was mentioned that when the Hausdorff dimension exceeds 3, self-intersections are inevitable (which is a rather trivial statement). We are on the opinion that it is necessary to clarify more precisely the occurrence of self-intersections in the helicalised fractals, by determining the upper bound to the Hausdorff dimension. This is because self-intersections would destroy the fractal-like structure of the infinitely helicalised curve, and a point of intersection is not well-defined since it corresponds to two different values in the domain. Therefore, we calculate the upper bounds to the Hausdorff dimension for the fractals of our interest as functions of the number of windings, with the details found in Section 5.

A strong motivation for having a comprehensive study on these helicalised curves and determining an upper bound to the Hausdorff dimension has to do with the structure of chromosomes. Through histones, a DNA double helix coils up to form nucleosomes, which would then further coil up many more times with the use of scaffold protein, forming levels upon levels of hyperhelices \cite{bio1,bio2,bio3, bio4}. According to Ref. \cite{bio1}, a human has enough DNA to go from the earth to the sun and back more than 300 times. This extensive length is wound up into the structure of an iteratively helicalised curve to form the chromosome, fitting snugly into the small nucleus of a cell. It is thus reasonable to use a helicalised fractal that is formulated here to approximate the structure of a chromosome. If the strand of DNA does not self-intersect, then the upper bound would provide a measure of how tight this packaging can possibly be.

In general, there have been some efforts to study various fractal properties associated to the structure of DNA \cite{DNA1,DNA2} as well as the structure of protein \cite{Prot1,Prot2,Prot3,Prot4,Prot5} - albeit not specifically dealing with how the structure of chromosomes enables DNA to be packed efficiently. Nevertheless, an intriguing idea was to think of the coding parts of a DNA sequence as clusters of connected sites in a random Cantor-like set (and conversely for the non-coding regions) \cite{DNA1}. Through their analysis, the resulting fractal dimension for lower organisms was 1, but of about 0.85 for higher eucaryotes - indicating a fractal coding/non-coding partitioning. This finding agreed with power law distribution of the non-coding parts of higher eucaryotes, with the pioneering work in such non-trivial clustering carried out by Ref. \cite{DNA2}. In addition, it is also interesting that the fractal dimension is related to the kinetic and thermodynamic aspects of proteins \cite{Prot2}, with other more recent developments in Ref. \cite{Prot3,Prot4,Prot5}.

\section{Sets of helicalisations}

We refer to the set of curves comprising the given smooth curve and all the subsequent helicalised curves as \emph{the set of helicalisations}. For instance if the starting curve is a straight line, then the set containing it with the helix, helicalised helix and so on is called \emph{the set of linear helicalisations} (since the starting curve is the ``linear'' straight line) (Fig. 1), with the infinitely helicalised line called the \emph{linear helical fractal}. We are also interested to study the case where the given curve is a circle, giving rise to \emph{the set of curved helicalisations} (since the starting curve is the ``curved'' circle) (Fig. 2, left column) and at the infinite level, the \emph{curved helical fractal}.

\begin{figure}[h]
\centering
\includegraphics[width=16cm]{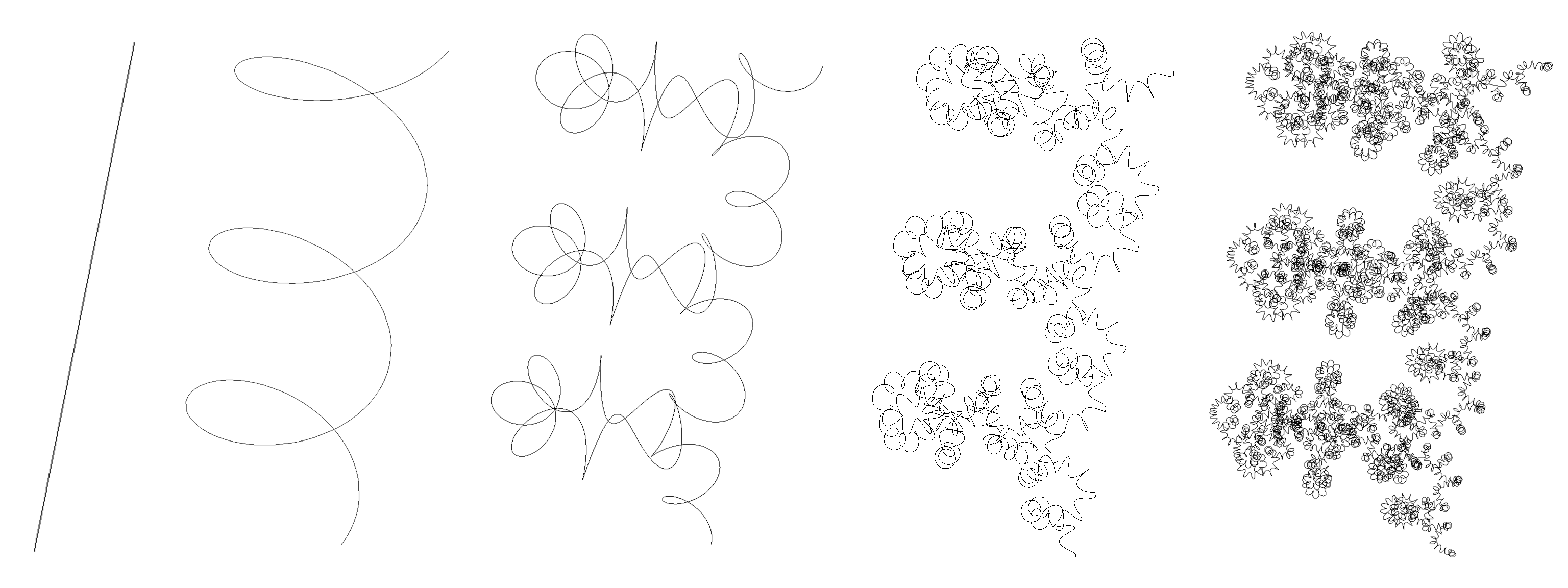}
\caption{The set of linear helicalisations, comprising the straight line, helix, helicalised helix, doubly helicalised helix, triply helicalised helix, and so on.}
\label{fig1}
\end{figure}

\begin{figure}
\centering
\includegraphics[width=13cm]{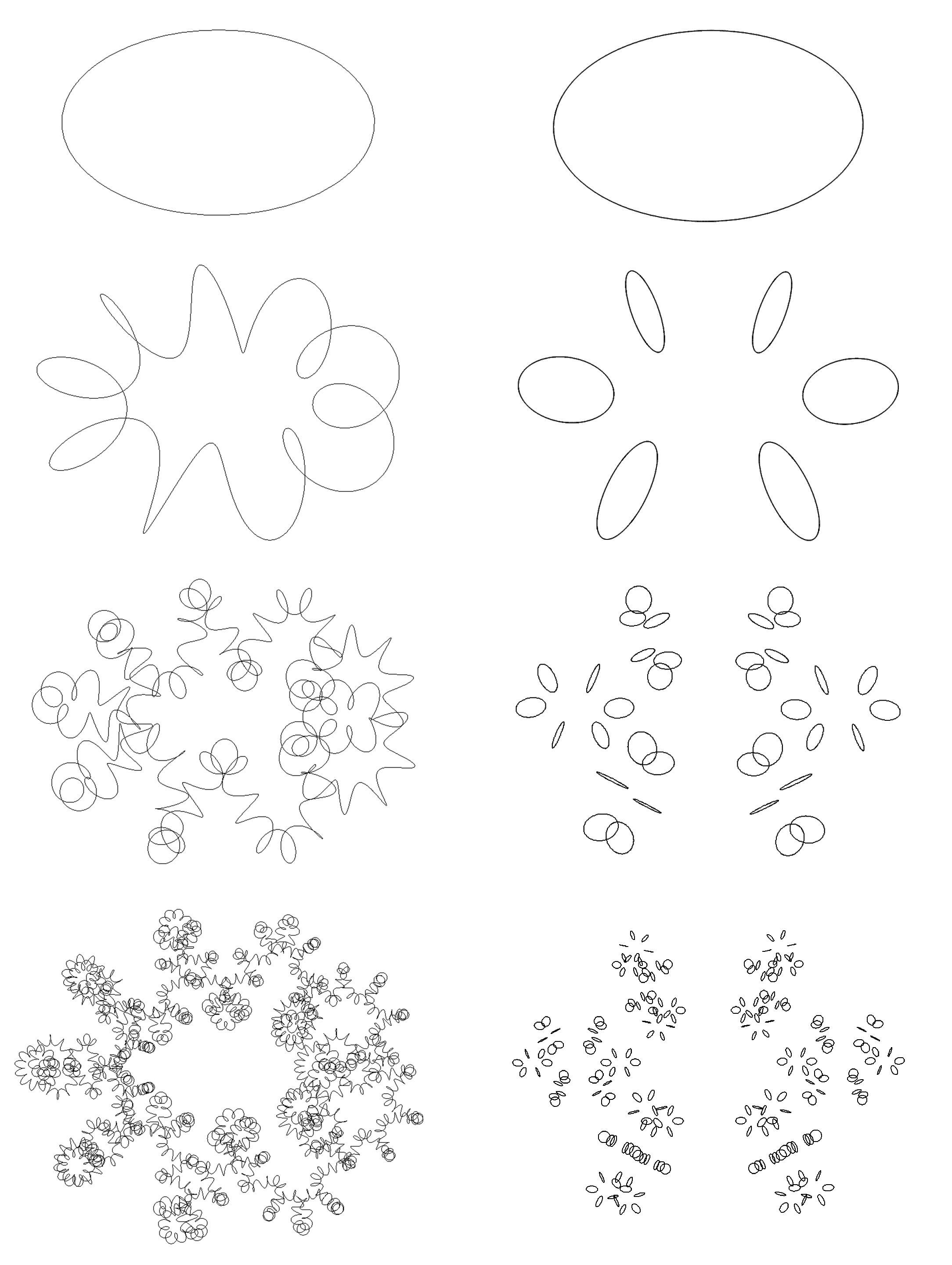}
\caption{Left column: The set of curved helicalisations, comprising the circle, toroidal curve, helicalised toroidal curve, doubly helicalised toroidal curve, and so on. Here, 9 new windings are added over each preceding winding, scaled by 1/3. Therefore, the Hausdorff dimension for the resulting curved helical fractal (see Eq. (\ref{D}) in Section 4) would be $D=2$. Right column: The set of discrete helicalisations, which is the discrete version of the set of curved helicalisations. Shown here are 6 new circles replacing each preceding circle, scaled by 1/3. The self-similar dimension of the resulting circular fractal is $D_S=1.63$, using Eq. (\ref{d}).}
\label{fig2}
\end{figure}

Before we formulate our definition of the helicaliser, it is perhaps insightful to consider the discrete version of this iterative helicalisation procedure applied to a circle of unit radius. Suppose $\omega=6$ and $R=1/3$. Then, the unit circle is replaced by six smaller circles with radius $R=1/3$ (Fig. 2, right column), each centred at the six equally spaced points on the circumference of the unit circle. Each such circle is orthogonal to the plane of the unit circle and lies in the plane containing the centre of the unit circle. It is clear that the resulting object is formed by six scaled copies of the given object. The next level would be to replace each of these six circles by another six circles (so the total number of circles will be thirty-six), whose radius is a third of that or 1/9 of the unit circle radius, arranged accordingly. From this observation, it may be concluded that in general since the $n$-th level is $\omega$ exact copies of the $(n-1)$-th level scaled down by $R$, the resulting object after infinitely many iterations is a fractal which we call the \emph{circular fractal}, with self-similar dimension \cite{text2}
\begin{eqnarray}\label{d}
D_S=-\frac{\ln{\omega}}{\ln{R}}.
\end{eqnarray}
We refer to the collection of the constituent levels as \emph{the set of discrete helicalisations}, since this is the discrete version of the curved helical fractal. Being self-similar, the circular fractal provides the crucial geometrical ingredients that motivate the formulation of the helicaliser. Suppose the starting curve is the unit circle. When the $n$-th level is helicalised to obtain the $(n+1)$-th level, it would have the following properties (abstracted from the discrete self-similar version):
\begin{enumerate}
\item
The windings must be added along directions that are everywhere orthogonal to the curve. This is justified since in the discrete version, the new circles are orthogonal to the old ones.
\item
The resulting $(n+1)$-th level has $\omega^n$ number of windings, each with radius $R^n$. This is based on the number of the new circles and the scaling, in the discrete version.
\end{enumerate}
With these properties, we will enunciate the precise definition of the helicaliser in the next section.

\section{Helicaliser: The procedure that replaces a smooth curve by one that winds around it}

Given a smooth curve $\vec{\psi}(v)=(\alpha(v),\beta(v),\gamma(v))$. The helicalised curve is defined as
\begin{eqnarray}\label{helicaliser}
\vec{\psi}_H(v)=\vec{\psi}(v)+R\cos{\omega v}\ \hat{n}_1(v)+R\sin{\omega v}\ \hat{n}_2(v),
\end{eqnarray}
where $0<R<1$ is the radius of the windings, and $\omega>1$ is an integer representing the number of windings of the helicalised curve around $\vec{\psi}$. The oscillatory terms $R\cos{\omega v}$ and $R\sin{\omega v}$ are added along a pair of orthonormal vectors $\hat{n}_1$ and $\hat{n}_2$, which are both orthonormal to the unit tangent vector $\hat{t}=\dot{\vec{\psi}}/|\dot{\vec{\psi}}|$ for all $v$. One may choose any set of functions $\hat{n}_1$ and $\hat{n}_2$ as long as they are smooth functions, with $\hat{t}$, $\hat{n}_1$, $\hat{n}_2$ forming a continuous triad of orthonormal vectors at all points along the given curve. One example would be the Frenet-Serret frame \cite{Pre} which is adopted in Ref. \cite{tolsua1} and involves the second derivative of $\vec{\psi}$. We shall be using a different set of orthonormal vectors instead, involving only the first derivative of $\vec{\psi}$. The unit tangent vector at any point is uniquely given by $\hat{t}=(\dot{\alpha},\dot{\beta},\dot{\gamma})/\sqrt{\dot{\alpha}^2+\dot{\beta}^2+\dot{\gamma}^2}$. Our choice is $\hat{n}_1=(-\dot{\beta},\dot{\alpha},0)/\sqrt{\dot{\alpha}^2+\dot{\beta}^2}$, with $\hat{n}_2=\hat{t}\times\hat{n}_1$.

Our interest in this study is mainly centered around the set of linear helicalisations and the set of curved helicalisations. For the former, the starting curve is the straight line $\vec{H}_1(v)=(0,0,v)$ with $v\in[0,2\pi]$. Applying the helicaliser as defined by Eq. (\ref{helicaliser}) to $\vec{H}_1$ yields the helix, though our prescription for $\hat{n}_1$ and $\hat{n}_2$ is undefined since $\alpha(v)=\beta(v)=0$. For this exceptional case, we let $\hat{n}_1=(1,0,0)$ and $\hat{n}_2=(0,1,0)$. The helix is then
\begin{eqnarray}\label{helix}
\vec{H}_2(v)=(R\cos{\omega v},R\sin{\omega v},v),
\end{eqnarray}
with $v\in[0,2\pi]$. Applying the helicaliser to $\vec{H}_2$ produces the helicalised helix. As mentioned in the previous section, each of the $\omega$ windings of the helix is to be replaced by $\omega$ windings whose radius is scaled by a factor of $R$. Consequently, the helicalised helix would have altogether $\omega^2$ windings with radius $R^2$. The parametric equations are
\begin{eqnarray}\label{helhelix}
\vec{H}_3(v)=
\left(
\begin{array}{c}
(1-R\cos{\omega^2v})R\cos{\omega v}\\
(1-R\cos{\omega^2v})R\sin{\omega v}\\
v\\
\end{array}
\right)
+\frac{R^2\sin{\omega^2v}}{\sqrt{1+\omega^2R^2}}
\left(
\begin{array}{c}
\sin{\omega v}\\
-\cos{\omega v}\\
\omega R\\
\end{array}
\right)
.
\end{eqnarray}
Subsequent levels are generated by iterative applications of the helicaliser where the domain of $v$ is always $[0,2\pi]$.

For the set of curved helicalisations, we take the starting curve to be the unit circle
\begin{eqnarray}\label{circle}
\vec{T}_1(v)=(\cos{v},\sin{v},0),
\end{eqnarray}
where $v\in[0,2\pi)$. Helicalising $\vec{T}_1$ produces the toroidal curve with parametric equations
\begin{eqnarray}\label{toroid}
\vec{T}_2(v)=
\left(
\begin{array}{c}
(1-R\cos{\omega v})\cos{v}\\
(1-R\cos{\omega v})\sin{v}\\
0\\
\end{array}
\right)
+\frac{R\sin{\omega v}}{\sqrt{2}}
\left(
\begin{array}{c}
\sin{v}\\
-\cos{v}\\
1\\
\end{array}
\right)
,
\end{eqnarray}
and so on.

\section{Hausdorff dimension: the relationship between the number of windings and the length of one winding as a cover}

The increase in the lengths of the elements of the set of helicalisations is an important property of the resulting fractal which can be used as a means of comparison amongst fractals with different $\omega$ and $R$. It is not easy however, to analytically compute these lengths, unlike the discrete case. For the set of linear helicalisations, although the lengths of the straight line and helix are easy to compute, lengths of the helicalised helix and higher levels are not computable without resorting to numerical methods. To keep our study analytically tractable, the lengths of the subsequent levels are approximated by unwrapping them into straight helices.

Once we have the total length of the $n$-th level, we can calculate the length of one winding for that level, and then use this as a cover of size $\varepsilon(n)$. Consequently, a total of $N(\varepsilon(n))=\omega^{n-1}$ covers of this size would be needed to cover the entire curve. The scaling exponent $D$ in the power law $N(\varepsilon(n))\sim1/\varepsilon(n)^D$ would be the Hausdorff dimension. Let us now proceed to find the length of the $n$-th level.

Let $L_n$ denote the length of $\vec{H}_n$, the $n$-th level of the set of linear helicalisations. From the previous section, the length of the straight line is simply $L_1=2\pi$. The length of the helix is
\begin{eqnarray}
L_2&=&\int_0^{2\pi}{\sqrt{\omega^2R^2\sin^2{\omega v}+\omega^2R^2\cos^2{\omega v}+1}}\ dv\\
&=&2\pi\sqrt{1+\omega^2R^2}.
\end{eqnarray}
For the helicalised helix, the length would be a complicated integral which is unlikely to be analytically integrable. Instead of turning to numerical computation, we model the helicalised helix as a straight helix with $\omega^2$ windings of radius $R^2$ equally spaced over the core length (length of straight axis) of $2\pi\sqrt{1+\omega^2R^2}$ as shown in Fig. 3. Such a helix would have parametric equations
\begin{eqnarray}
\vec{H}^*_3(v)=\left(R^2\cos{\left(\frac{\omega^2}{\sqrt{1+\omega^2R^2}}v\right)},R^2\sin{\left(\frac{\omega^2}{\sqrt{1+\omega^2R^2}}v\right)},v\right),
\end{eqnarray}
with $v\in[0,2\pi\sqrt{1+\omega^2R^2}]$. Thus, the length of the helicalised helix under this approximation is
\begin{eqnarray}
L_3&=&\int_0^{2\pi\sqrt{1+\omega^2R^2}}{\sqrt{\frac{\omega^4R^4}{1+\omega^2R^2}\left[\sin^2{\left(\frac{\omega^2}{\sqrt{1+R^2\omega^2}}v\right)}+\cos^2{\left(\frac{\omega^2}{\sqrt{1+R^2\omega^2}}v\right)}\right]+1}}\ dv\ \\
&=&2\pi\sqrt{1+\omega^2R^2+\omega^4R^4}.
\end{eqnarray}

\begin{figure}
\centering
\includegraphics[width=16cm]{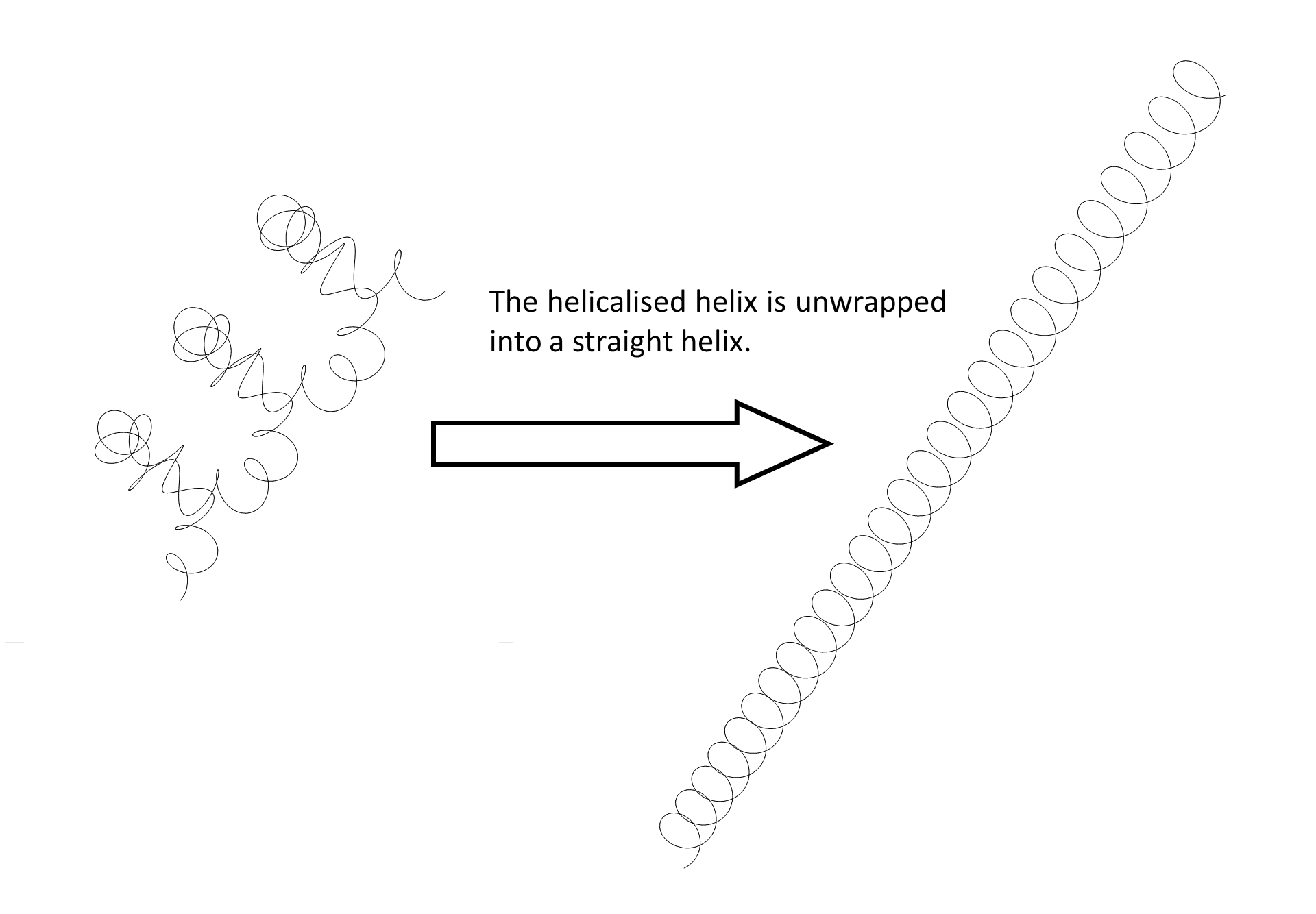}
\caption{The length of the helicalised helix is approximated by unwrapping it into a straight helix.}
\label{fig3}
\end{figure}

In general, the length of the $n$-th level of the set of linear helicalisations under this approximation is given by
\begin{eqnarray}\label{Ln}
L_n=2\pi\sqrt{\sum_{i=0}^{n-1}{(\omega R)^{2i}}}=2\pi\sqrt{\frac{1-(\omega R)^{2n}}{1-(\omega R)^2}}.
\end{eqnarray}
To prove this, the parametric equations for $\vec{H}^*_{n+1}$ is derived under the assumption that $L_n$ is given by Eq. (\ref{Ln}). Here, $\vec{H}^*_{n+1}$ is a straight helix with core length $L_n$, and has $\omega^n$ windings of radius $R^n$, so
\begin{eqnarray}
\vec{H}^*_{n+1}(v)=\left(R^n\cos{\left(\frac{2\pi\omega^n}{L_n}v\right)},R^n\sin{\left(\frac{2\pi\omega^n}{L_n}v\right)},v\right),
\end{eqnarray}
where $v\in[0,L_n]$. Therefore,
\begin{eqnarray}
L_{n+1}&=&\int_0^{L_n}{\sqrt{\left(\frac{2\pi}{L_n}\right)^2(\omega R)^{2n}+1}}\ dv\\
&=&\sqrt{(2\pi)^2(\omega R)^{2n}+L_n^2}\\
&=&2\pi\sqrt{(\omega R)^{2n}+\frac{1-(\omega R)^{2n}}{1-(\omega R)^{2}}}\\
&=&2\pi\sqrt{\frac{1-(\omega R)^{2(n+1)}}{1-(\omega R)^{2}}},
\end{eqnarray}
thereby completing the proof by induction since Eq. (\ref{Ln}) is valid for $n=1$ that gives $L_1=2\pi$.

For the set of curved helicalisations, the length of the unit circle $\vec{T}_1$ is $M_1=2\pi$. Just like the set of linear helicalisations, the length of the toroidal curve is approximated by detaching (removing the point where $v=0$ for instance) and then unwrapping it into a straight helix of core length $M_1$ with $\omega$ windings of radius $R$. Therefore, its parametric equations are
\begin{eqnarray}
\vec{T}^*_2(v)=\left(R\cos{\omega v},R\sin{\omega v},v\right),
\end{eqnarray}
with $v\in(0,2\pi)$. This then gives $M_2=2\pi\sqrt{1+\omega^2R^2}$. It should be clear that subsequent levels for the set of curved helicalisations all have the same approximate lengths as the corresponding levels for the set of linear helicalisations, so that $M_n=L_n$.

The $n$-th level of both the sets of linear helicalisations and curved helicalisations has $\omega^{n-1}$ windings (with the exception of $n=1$), with total length $L_n$. Consequently, each of these windings would have a length of $L_n/\omega^{n-1}$. The Hausdorff dimension is then
\begin{eqnarray}\label{defHaus}
D=-\lim_{n\rightarrow\infty}\frac{\ln{\omega^{n-1}}}{\ln{(L_n/\omega^{n-1})}},
\end{eqnarray}
since at the $n$-th level, a total of $\omega^{n-1}$ windings being covers with size $L_n/\omega^{n-1}$ would be required to cover the entire curve. Evaluating the limits in the respective domains (see appendix A for the explicit calculations):

For $\omega R\leq1$,
\begin{eqnarray}\label{Dsmallerthanone}
D&=&1,
\end{eqnarray}
whereas for $\omega R>1$,
\begin{eqnarray}\label{D}
D&=&-\frac{\ln{\omega}}{\ln{R}}.
\end{eqnarray}
These results also show that the Hausdorff dimension $D(\omega, R)$ is everywhere continuous.

Incidentally, the self-similar dimension of a self-similar fractal, Eq. (\ref{d}), makes use of the scale factor $k$, which would here correspond to the ratio of the length of one winding in the $n$-th level to that in the $(n+1)$-th level. But since these helicalised fractals are not exactly self-similar, the ramification is that $k(n)$ is not a constant:
\begin{eqnarray}\label{kn}
k(n)=\omega\sqrt{\frac{1-(\omega R)^{2n}}{1-(\omega R)^{2n+2}}}.
\label{sfactor}
\end{eqnarray}
Nevertheless, it is intriguing to observe that if we take the limit of $k(n)$ as $n$ goes to infinity, we recover the same results as the Hausdorff dimension Eqs. (\ref{Dsmallerthanone}) and (\ref{D}) or the expression for the self-similar dimension Eq. (\ref{d}) of a self-similar fractal with lower bound 1 (see appendix B for the explicit calculations):

For $\omega R\leq1$,
\begin{eqnarray}\label{D'smallerthanone}
D'&=&1,
\end{eqnarray}
whereas for $\omega R>1$,
\begin{eqnarray}\label{D'}
D'&=&-\frac{\ln{\omega}}{\ln{R}},
\end{eqnarray}
where $D'$ is defined as $\displaystyle D'=\lim_{n\rightarrow\infty}\frac{\ln{\omega}}{\ln{k(n)}}$. Observe also that for the set of discrete helicalisations, since the scale factor would be a constant $k=1/R$ independent of $n$, Eq. (\ref{D'}) is simply the self-similar dimension for the self-similar circular fractal for all values of $\omega$ and $R$, with no lower bound.

These results say that the resulting linear helical fractal and curved helical fractal have Hausdorff dimension $D=1$ with $\displaystyle\lim_{n\rightarrow\infty}{L_n}$ being finite whenever $\omega R<1$, just like any finite length curve is a one-dimensional object. The case $\omega R=1$ is when $\displaystyle\lim_{n\rightarrow\infty}{L_n}$ just becomes infinite, with $D=1$ as well. When $\omega R>1$, the fractals take on a non-integer Hausdorff dimension $D>1$ with $\displaystyle\lim_{n\rightarrow\infty}{L_n}=\infty$. So the fractals have a lower bound to $D$, which is 1. This is unlike the discrete version in the case of the circular fractal, where the self-similar dimension is given by Eq. (\ref{d}) for all values of $\omega$ and $R$. The length of the infinite level is zero when $\omega R<1$, corresponding to $D_S<1$, and the length is finite ($2\pi$, equal to the circumference of the starting unit circle) when $\omega R=1$, corresponding to $D_S=1$. The self-similar dimension is greater than 1 when $\omega R>1$, with the resulting fractal having infinite length.

The reason for this difference between the helicalised curve and the discrete version is that in the latter, a circle is replaced by discrete circles whose total length can be smaller than the one it replaces if $\omega R<1$. Repetition of this process would produce a collection of circles whose total length diminishes. For the former on the other hand, the helicalised curve must have a length that is necessarily longer because the windings around the core curve are connected. The calculations show that the increase in length of a curve helicalised infinitely leads to a finite length for $\omega R<1$, with the length growing unbounded otherwise.

\section{Upper bounds to the Hausdorff dimension for the curved helical fractal and linear helical fractal}

We have shown in the previous section that the Hausdorff dimension for the linear helical fractal and the curved helical fractal must be at least $D=1$, and it takes the same form as the self-similar dimension of a self-similar fractal $D_S$ in Eq. (\ref{d}) for $\omega R>1$. Here, we will derive the upper bounds based on the constraint that the resulting infinitely helicalised fractal must not self-intersect. This geometrical constraint is crucial, because self-intersection makes the object lose its fractal-like properties. We first focus on the discrete self-similar circular fractal to calculate the exact upper bound to $D_S$, and this would be an upper bound to $D$ for the curved helical fractal. An upper bound to $D$ for the linear helical fractal is stated right after, with its corresponding explicit derivation given in appendix C.

Fig. 4 shows the angled and top views of the first four levels of the set of discrete helicalisations that are superimposed, where $\omega=6$ and $R=1/3$. The corresponding figure for the set of curved helicalisations is shown in Fig. 5. Self-intersections would occur if there are too many new circles or windings introduced in the succeeding levels that are not sufficiently scaled down. Fig. 6 shows how this crowding of the circles happens for the set of discrete helicalisations. The six level two circles which are arranged along the unit circle would be most crowded along the inner parts, where subsequent levels are filled in. As the levels increase, the points labeled $A$ to $F$ would be surrounded by more and more circles that increasingly take up more space. The dotted circles around these points indicate the ultimate regions that are occupied by those particular circles of the infinite level. If the radius of these dotted circles is denoted as $r_\infty$, then
\begin{eqnarray}\label{rinfty}
r_\infty=\sum_{i=2}^{\infty}{R^i}=\frac{R^2}{1-R}.
\end{eqnarray}
A similar crowding happens for the set of curved helicalisations, with the curved helical fractal having a crowding at the inner region of its level one circle. As can be observed in Fig. 5, the process of constructing the curved helical fractal would result in the radii of windings of levels three, four, and so on, adding up to the crowding. This ultimate crowding is similar to the discrete case, i.e. having the six dotted circles with radius $r_\infty$ in Eq. (\ref{rinfty}), although these dotted circles acquire a tilt. This requires some visualisation to understand how the windings are oriented (and subsequently how the crowding occurs) since in the case of the curved helical fractal, the level two toroidal curve is tilted with respect to the level one circle as opposed to being perpendicular. In our calculations that follow, we will ignore the tilt and treat it identically as the case of the self-similar circular fractal (as the centres of these dotted circles all lie in the same plane, just like the discrete case), which would considerably simplify the analysis for deriving an upper bound to $D$ for the curved helical fractal. To avoid having to use two different symbols, we will refer to both the upper bound to the self-similar dimension of the circular fractal and an upper bound to the Hausdorff dimension of the curved helical fractal as $D_\textrm{upper}$.

\begin{figure}
\centering
\includegraphics[width=16cm]{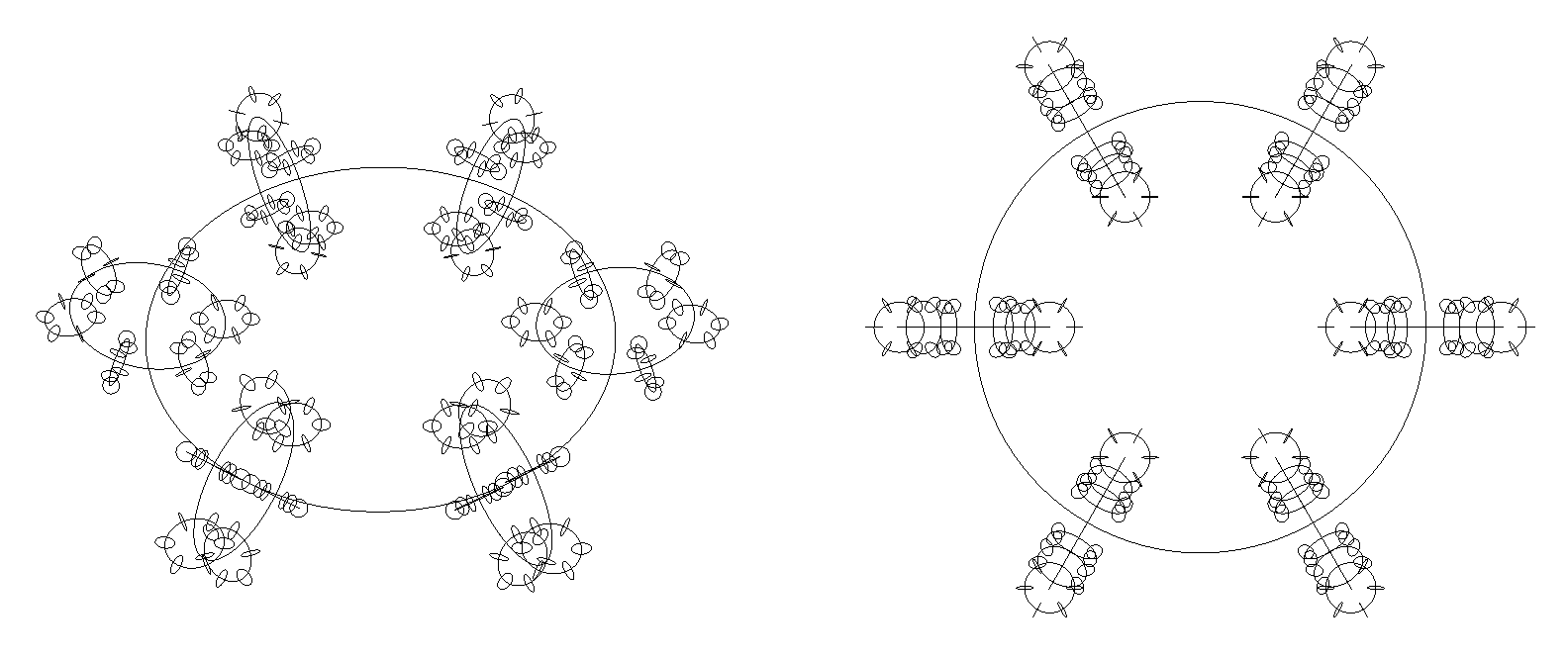}
\caption{Angled and top views of the first four levels of the set of discrete helicalisations superimposed.}
\label{fig4}
\end{figure}

\begin{figure}
\centering
\includegraphics[width=16cm]{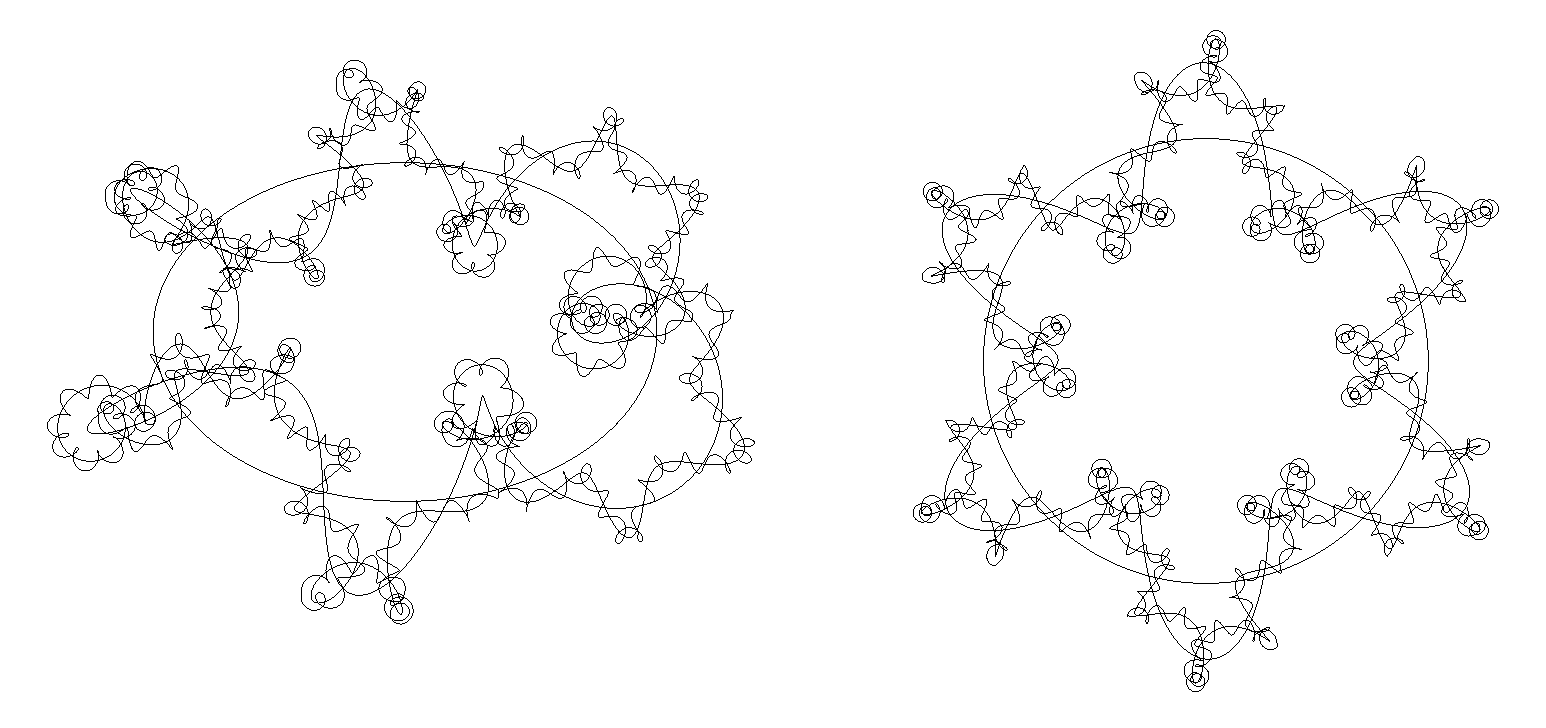}
\caption{Angled and top views of the first four levels of the set of curved helicalisations superimposed.}
\label{fig5}
\end{figure}

\begin{figure}
\centering
\includegraphics[width=12cm]{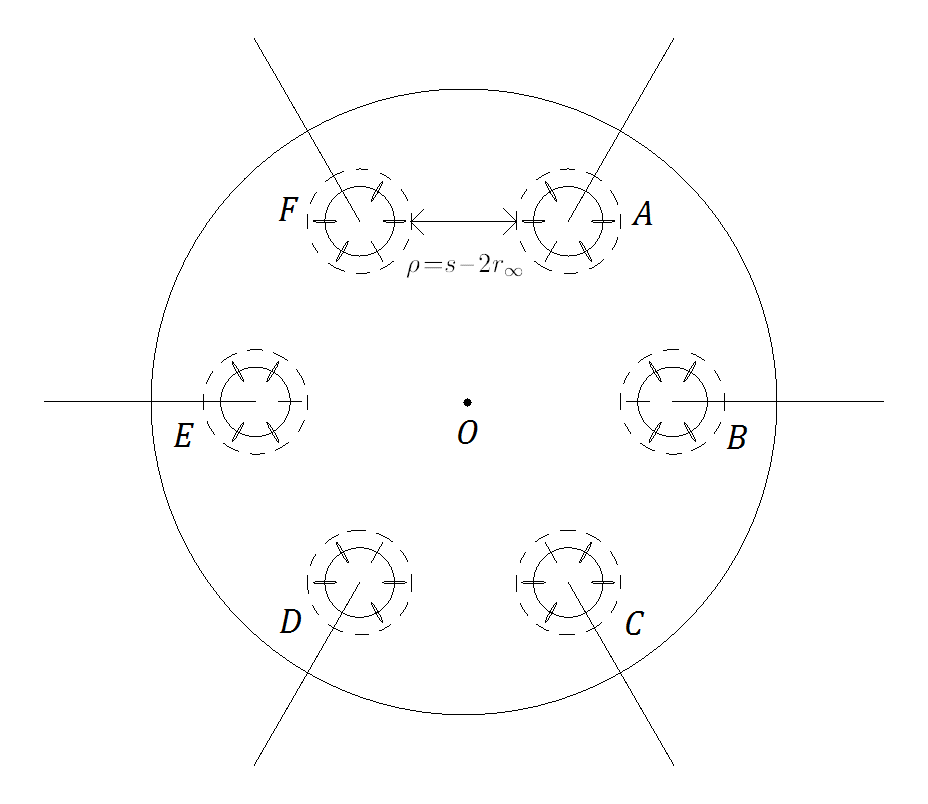}
\caption{Crowding of the circles of the set of discrete helicalisations at the inner region. The points $A$ to $F$ are centres of the dotted circles. The curved helical fractal would be treated to have the same effective crowding, represented by the same dotted circles of the same radius, $r_\infty$.}
\label{fig6}
\end{figure}

Let the distance between two adjacent centres of these dotted circles around $A$ to $F$ be $s$, and the shortest distance between two such adjacent dotted circles be $\rho$. The condition that self-intersections do not occur would be
\begin{eqnarray}\label{rho}
\rho=s-2r_\infty>0.
\end{eqnarray}
The distance $s$ can be found as follows. The points $A$ to $F$ form a regular hexagon with centre $O$. For a general situation with $\omega$ circles (or windings), these form a regular $\omega$-gon so that $\angle AOB=2\pi/\omega$. Also, $OA=OB=1-R$ and hence $s=2\mu(1-R)$, where $\mu=\sin{(\pi/\omega)}$. Thus,
\begin{eqnarray}\label{rhooo}
\rho=2\mu(1-R)-\frac{2R^2}{1-R}.
\end{eqnarray}
Solving for $R$ gives
\begin{eqnarray}
R=\frac{\rho-4\mu\pm\sqrt{\rho^2-8\rho+16\mu}}{4(1-\mu)}.
\end{eqnarray}
To see which sign on the root to take, note that $\rho>0$. In the limiting situation where $\rho$ approaches zero in that the circles are almost touching,
\begin{eqnarray}
\lim_{\rho\rightarrow0^+}{R}=\frac{-\mu\pm\sqrt{\mu}}{1-\mu}.
\end{eqnarray}
Now $0<\mu\leq1$, so the denominator is non-negative and also $0<\mu\leq\sqrt{\mu}\leq1$. As a result, taking the positive root gives $R\geq0$, whilst taking the negative root yields $R<0$. The latter is thus invalid and the positive root is taken. In addition, $\sqrt{\mu}\leq1$ implies that $\sqrt{\mu}-\mu\leq1-\mu$. In fact, $\mu$ is strictly less than 1 when $\omega=3,4,5,...$ . Under these conditions therefore, $0<R<1$ which is consistent with the basic requirements. Thus,
\begin{eqnarray}
R=\frac{\rho-4\mu+\sqrt{\rho^2-8\rho+16\mu}}{4(1-\mu)}.
\end{eqnarray}
Using Eq. (\ref{D}) to eliminate $R$,
\begin{eqnarray}\label{Dupper}
D=\frac{\ln{\omega}}{\ln{4}+\ln{(1-\mu)-\ln{(\rho-4\mu+\sqrt{\rho^2-8\rho+16\mu})}}}.
\end{eqnarray}
The value of $\rho$ is continuous since the distance between two adjacent dotted circles is geometrically continuous. This implies that $D$ is also continuous despite the fact that $\omega$ is a discrete integer.

The no-self-intersection constraint given in Eq. (\ref{rho}) that prevents the destruction of the fractal structure would place an upper bound to $D$. In the limiting case where $\rho=0$, Eq. (\ref{Dupper}) reduces to
\begin{eqnarray}\label{Dupperrr}
D_{\textrm{upper}}(\omega)=\frac{\ln{\omega}}{\ln{(1+\sqrt{\mu})}-\ln{\sqrt{\mu}}}.
\end{eqnarray}
Notice that $D_{\textrm{upper}}$ is discrete, depending on the discrete number of windings or new circles being used. Two plots of $D_{\textrm{upper}}$ against $\omega$ for different domains of $\omega$ are shown in Fig. 7. For $\omega=6$, $D_{\textrm{upper}}=2.03292$. The general variation of $D_{\textrm{upper}}$ is that it increases from 1.50546 at $\omega=3$ and achieves maximum value of 2.43589 at $\omega=67$. For larger values of $\omega$, $D_{\textrm{upper}}$ then decreases monotonically approaching 2 as $\omega$ goes to infinity. This suggests that for a very large number of windings (where the windings are so tight, resembling a 2-d tube), the curved helical fractal and circular fractal would lose their fractal structure due to self-intersection if they take up more ``space'' than a two-dimensional surface would. The densest of such fractals (where $D$ is very close to $D_\textrm{upper}$) with an enormous number of windings may share similar properties to those of an ordinary surface and one may for instance define an ``area'' for these fractals, though we shall not pursue further on these properties here in this paper.

\begin{figure}
\centering
\includegraphics[width=16cm]{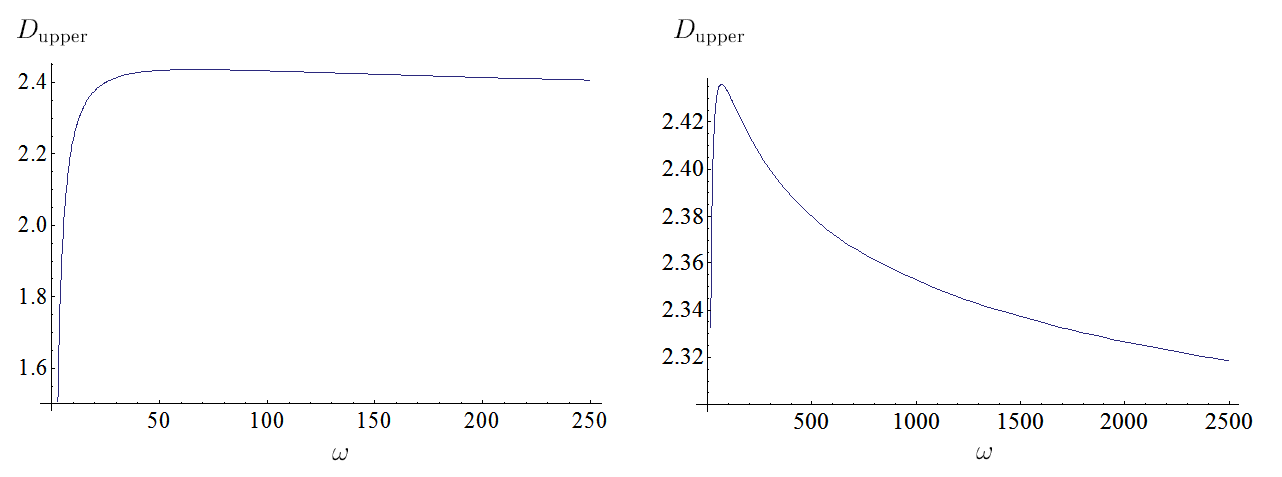}
\caption{Two plots of $D_\textrm{upper}$ for the curved helical fractal and circular fractal against $\omega$ for different domains of $\omega$. Note that $D_\textrm{upper}$ increases from 1.50546 at $\omega=3$ and achieves maximum value of 2.43589 at $\omega=67$. For larger values of $\omega$, $D_\textrm{upper}$ decreases monotonically to 2 as $\omega\rightarrow\infty$.}
\label{fig7}
\end{figure}

\begin{figure}
\centering
\includegraphics[width=16cm]{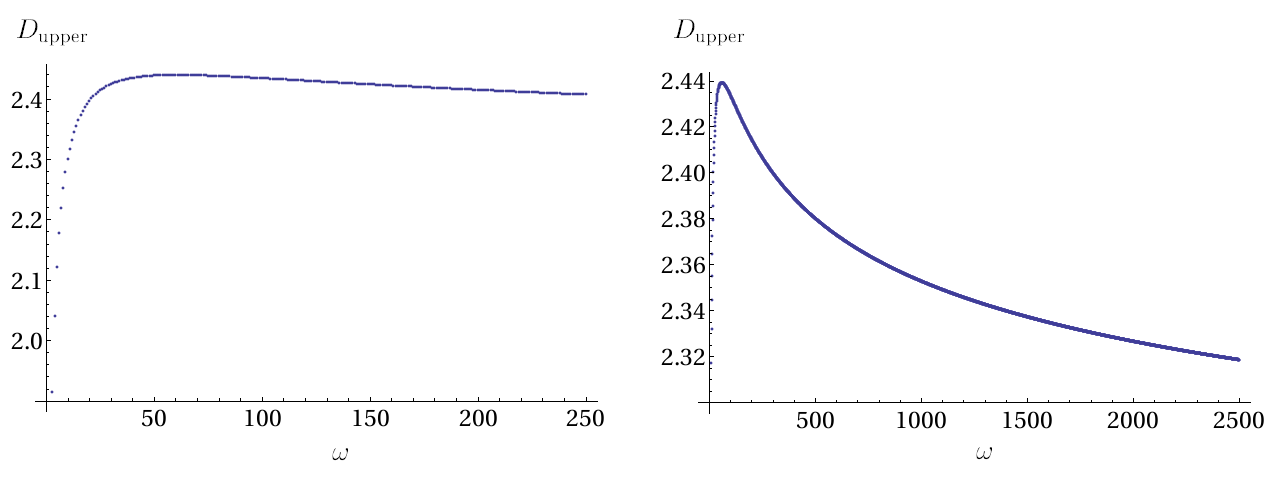}
\caption{The corresponding plots of $D_\textrm{upper}$ against $\omega$ for the linear helical fractal. These plots are obtained by solving for $R$ in Eq. (\ref{Dupperhelix}) numerically given integer values of $\omega$, and then calculating $D_{\textrm{upper}}$ using Eq. (\ref{D}). For every $\omega$, the value of $D_\textrm{upper}$ for the linear helical fractal is always larger than the corresponding one for the curved helical fractal due to the extra space from the displacement along the direction of $\vec{H}_1$. The properties of $D_\textrm{upper}$ against $\omega$ for the linear helical fractal are similar to those of the curved helical fractal.}
\label{fig8}
\end{figure}

The corresponding plots for the linear helical fractal are shown in Fig. 8, with the detailed derivation presented in the appendix C. Although not identical to the case for the circular fractal, they do share the same essential features. In particular, the limit as $\omega\rightarrow\infty$ is also 2.

\section{Concluding remarks}

In this paper, we have gone through a systematic study on the helicaliser, whereby the manner for a curve being iteratively helicalised to form the fractal (the $n$-th level is helicalised by adding $\omega^n$ new windings scaled by $R^n$ along directions orthogonal to the tangent vector of the curve) is based on the properties of the discrete self-similar circular fractal. This geometrically motivated framework therefore provides a platform which can be used for future development on this topic. Our investigations show that the self-similar discrete version serves as a fundamental comparison for the helicalised counterpart.

It is intriguing that for very large $\omega$, the upper bound $D_{\textrm{upper}}$ has the limit of 2, implying that the coiling structure of very tightly packed DNA in a chromosome resembles a two-dimensional surface. This would suggest that a topological analysis on this biological structure by modelling it as an ordinary surface may be beneficial in obtaining new insights towards understanding the behavior and dynamics of chromosomes. For instance, certain topological invariants for 2-manifolds \cite{topo1,topo2} or ideas derived from knot theory \cite{knot1,knot2,knot3} may give rise to some constraints which turn out to explain the properties of DNA packaging mechanism. We emphasise that our result in Section 5 indicates that in order to accurately model it as a surface, one should consider very large $\omega$ because it is in that limit that the Hausdorff dimension approaches 2. Otherwise, the DNA structure may not correctly be treated as a surface since the upper bound would usually be greater than 2 (sometimes 2.4 for certain values of $\omega$). 

As a final remark, we would like to mention that the idea of helicalising a curve has inspired a general method of constructing manifolds of revolution around a given curve, which represent 4-d (or any $n$-d) spacetime. Curved traversable wormholes, for example the helical wormhole and the catenary wormhole, are constructed by this modified helicalisation method (to produce manifolds instead of curves around a given curve) which have the desired property of containing safe geodesics (locally supported by non-exotic matter) through the wormholes \cite{Vee2012,Vee2013}. This general method of constructing manifolds of revolution around a given curve is also used to construct spacetimes composed of rotating shells \cite{Vee2014}. In doing so, Ref. \cite{Vee2014} provides a geometrical visualisation that describes how the various coordinate systems for the Schwarzschild metric arise from the freedom in parametrising the straight line and the radial function from this general method.

\appendix
\section{Explicit calculations of the Hausdorff dimension for $\omega R<1$, $\omega R=1$ and $\omega R>1$}

For $\omega R<1$, $L_n=2\pi\sqrt{(1-(\omega R)^{2n})/(1-(\omega R)^2)}$ which is approximately $2\pi/\sqrt{1-(\omega R)^2}$ for large $n$. Then from Eq. (\ref{defHaus}),
\begin{eqnarray}
D&=&-\lim_{n\rightarrow\infty}{\frac{\ln{\omega^{n-1}}}{\ln{(L_n/\omega^{n-1})}}}\\
&=&-\lim_{n\rightarrow\infty}{\frac{(n-1)\ln{\omega}}{\ln{\left(2\pi/\sqrt{1-(\omega R)^2}\right)}-(n-1)\ln{\omega}}}\\
&=&-\lim_{n\rightarrow\infty}{\frac{(n-1)\ln{\omega}}{-(n-1)\ln{\omega}}}\\
&=&1.
\end{eqnarray}
For $\omega R>1$, $L_n=2\pi\sqrt{(1-(\omega R)^{2n})/(1-(\omega R)^2)}$ which is approximately $2\pi(\omega R)^n/\sqrt{(\omega R)^2-1}$ for large $n$. Then 
\begin{eqnarray}
D&=&-\lim_{n\rightarrow\infty}{\frac{\ln{\omega^{n-1}}}{\ln{(L_n/\omega^{n-1})}}}\\
&=&-\lim_{n\rightarrow\infty}{\frac{(n-1)\ln{\omega}}{\ln{\left(2\pi/\sqrt{(\omega R)^2-1}\right)}+\ln{\omega}+n\ln{R}}}\\
&=&-\lim_{n\rightarrow\infty}{\left(\frac{n-1}{n}\right)\frac{\ln{\omega}}{\ln{R}}}\\
&=&-\frac{\ln{\omega}}{\ln{R}}.
\end{eqnarray}
For $\omega R=1$, note that $L_n$ from Eq. (\ref{Ln}) gives $L_n=2\pi\sqrt{n}$, so that the length of one winding of the $n$-th level is $2\pi\sqrt{n}/\omega^{n-1}$. Then
\begin{eqnarray}
D&=&-\lim_{n\rightarrow\infty}{\frac{\ln{\omega^{n-1}}}{\ln{(L_n/\omega^{n-1})}}}\\
&=&-\lim_{n\rightarrow\infty}\frac{(n-1)\ln{\omega}}{\ln{(2\pi)+0.5\ln{n}-(n-1)\ln{\omega}}}\\
&=&-\lim_{n\rightarrow\infty}\frac{(n-1)\ln{\omega}}{-(n-1)\ln{\omega}}\\
&=&1.
\end{eqnarray}

Summarising,
\begin{eqnarray}
D&=&1,\textrm{\ for\ }\omega R\leq1,\\
D&=&-\frac{\ln{\omega}}{\ln{R}},\text{\ for\ }\omega R>1,
\end{eqnarray}
which are Eqs. (\ref{Dsmallerthanone}) and (\ref{D}).

\section{Explicit calculations of limit of $k(n)$ as $n\rightarrow\infty$ for $\omega R<1$, $\omega R=1$ and $\omega R>1$}

The scale factor $k(n)$ is given by Eq. (\ref{kn}). For $\omega R<1$,
\begin{eqnarray}
\lim_{n\rightarrow\infty}{k(n)}=\omega\sqrt{\frac{1-0}{1-0}}=\omega.
\end{eqnarray} 
For $\omega R>1$,
\begin{eqnarray}
\lim_{n\rightarrow\infty}{k(n)}=\omega\sqrt{\frac{1/(\omega R)^{2n}-1}{1/(\omega R)^{2n}-(\omega R)^2}}=\frac{1}{R}.
\end{eqnarray}
For $\omega R=1$, note that $L_n$ from Eq. (\ref{Ln}) gives $L_n=2\pi\sqrt{n}$, so that the length of one winding of the $n$-th level is $2\pi\sqrt{n}/\omega^{n-1}$. This gives $k(n)=\omega\sqrt{n/(n+1)}$. Therefore,
\begin{eqnarray}
\lim_{n\rightarrow\infty}{k(n)}=\omega.
\end{eqnarray}
Summarising,
\begin{eqnarray}
D'&=&1,\textrm{\ for\ }\omega R\leq1,\\
D'&=&-\frac{\ln{\omega}}{\ln{R}},\text{\ for\ }\omega R>1,
\end{eqnarray}
which are Eqs. (\ref{D'smallerthanone}) and (\ref{D'}).

\section{Derivation of an upper bound to the Hausdorff dimension for the linear helical fractal}

For the linear helical fractal, apart from the crowding of the dotted circles (see Fig. 6) $A$ to $F$ in the transverse plane (this plane is perpendicular to the level one straight line, $\vec{H}_1$), these dotted circles are displaced along the direction of $\vec{H}_1$ and then tilted since the oscillatory terms are added orthonormal to the curve. The discrete version of the set of linear helicalisations is shown in Fig. 9 to assist in visualisation. When $v$ runs from 0 to $2\pi$, this displacement of $2\pi$ along $\vec{H}_1$ is shared by $\omega$ windings, so each winding gets a displacement of $2\pi/\omega$. Therefore, the displacement along $\vec{H}_1$ between the two adjacent dotted circles is $2\pi/\omega^2$. The upper bound to $D$ is reached when these displaced and tilted adjacent dotted circles are touching, so that from the Pythagorean theorem,
\begin{eqnarray}\label{preDupperhelix}
\left(2r'_\infty\right)^2=s'^2+\left(\frac{2\pi}{\omega^2}\right)^2,
\end{eqnarray}
where $r'_\infty=\displaystyle\sum_{i=3}^\infty{R^i}=\frac{R^3}{1-R}$ and $s'=2\mu R(1-R)$, recalling that $\mu=\sin{\left(\pi/\omega\right)}$. (The primes here and in the next subsection do not represent derivatives. Instead, they denote new variables.) The extra factor of $R$ here is because the helix has radius $R$ while the circle (in the construction for the curved helical fractal and circular fractal) has unit radius. Some algebraic manipulations would lead to
\begin{eqnarray}\label{Dupperhelix}
R^6=\mu^2R^2(1-R)^4+\frac{\pi^2}{\omega^4}(1-R)^2.
\end{eqnarray}

\begin{figure}
\centering
\includegraphics[width=16cm]{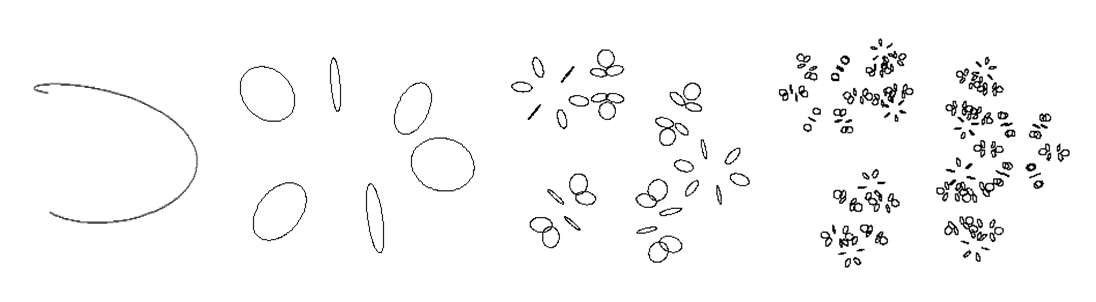}
\caption{The discrete version of the set of linear helicalisations, where one winding of the helix is shown here. In comparison with Fig. 2 for the sets of curved and discrete helicalisations, the six circles replacing the helix are displaced along $\vec{H}_1$ which yields more space between them than those for the curved helical fractal and circular fractal.}
\label{fig9}
\end{figure}

The additional term $\pi^2(1-R)^2/\omega^4$ due to the displacement along $\vec{H}_1$ (as compared to Eq. (\ref{rhooo}) with $\rho=0$) results in $D_{\textrm{upper}}$ for the linear helical fractal being larger than that for the curved helical fractal and circular fractal, since this displacement gives more space and hence allows a greater number of windings or windings with larger radius. The presence of this term makes it difficult to solve for $R$ analytically. We have not found the closed form of $R$, nor the analogue to Eq. (\ref{Dupperrr}). Nevertheless, $D_{\textrm{upper}}$ can be numerically solved with the graphs shown in Fig. 8. The general features for $D_{\textrm{upper}}$ of the linear helical fractal are similar to those of the curved helical fractal and circular fractal, viz. that it increases from 1.91349 at $\omega=3$ and achieves maximum value of 2.43913 at $\omega=61$, with $D_{\textrm{upper}}$ decreasing monotonically to 2 as $\omega$ grows to infinity. These comparable properties between $D_{\textrm{upper}}$ of the fractals are due to the fact that as $\omega$ gets larger, the displacement along $\vec{H}_1$ for the linear helical fractal gets reciprocally smaller (since that displacement is $2\pi/\omega^2$), which reduces to the case of the curved helical fractal.

\subsection{Self-intersections do not occur at finer scales, given these upper bounds}

Consider again the circular fractal. The calculations that we did for $D_\textrm{upper}$ were carried out by looking at the crowding at the inner region of the level one circle. What would the crowding be if it is seen from the point of view of a level two circle? As can be seen in Fig. 10, since the circular fractal is self-similar, the crowding at the inner region of a level two circle is identical to that for the level one circle. The same would be true from the perspectives of higher levels (finer and even finer scales), which is the consequence of its self-similarity. Therefore, $D_\textrm{upper}$ is exactly the one given by Eq. (\ref{Dupperrr}).

\begin{figure}[h]
\centering
\includegraphics[width=16cm]{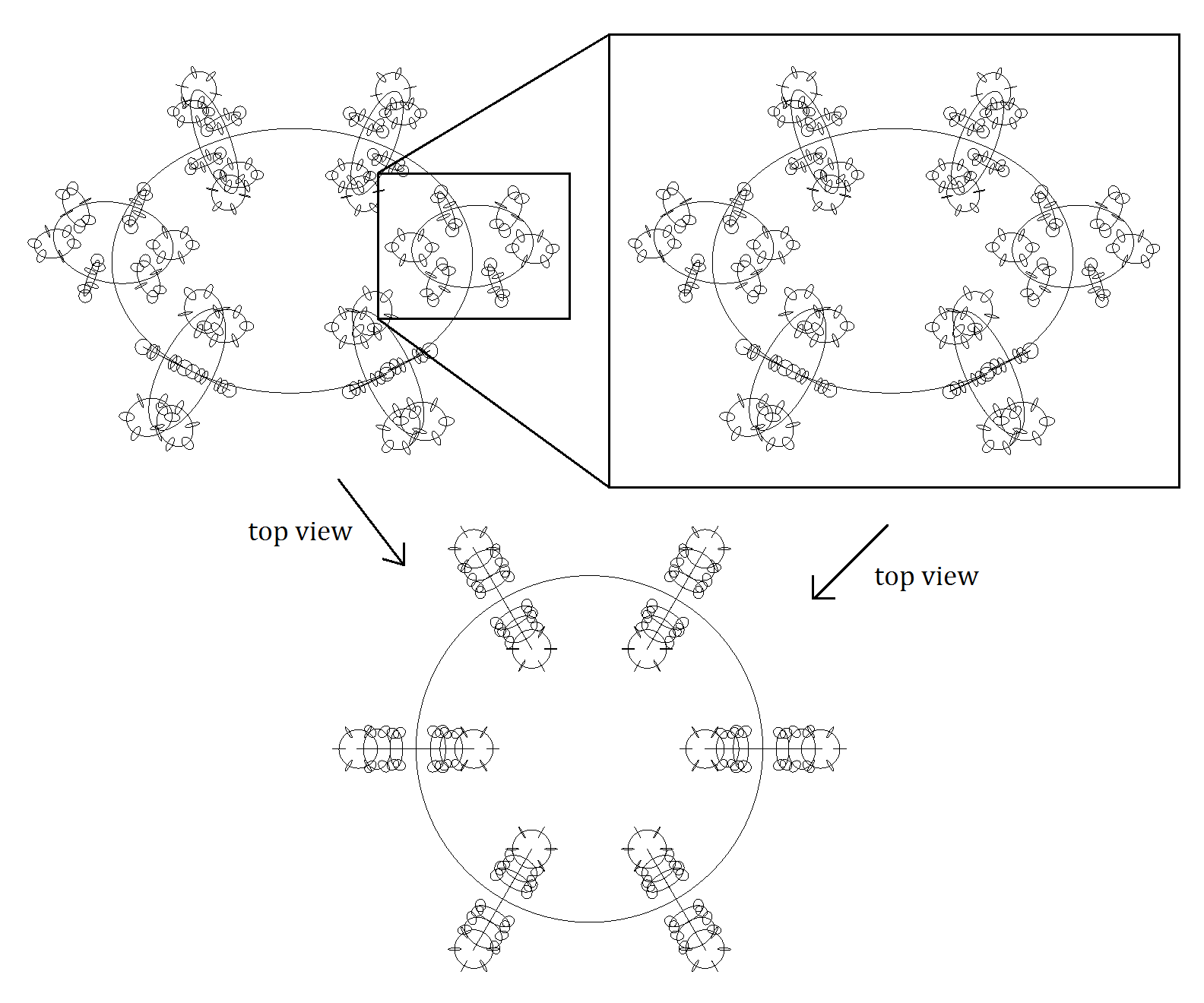}
\caption{The crowding of the circular fractal as seen from the point of view of the level one circle and from one of its level two circle. Due to its self-similarity, this view is the same for all scales. For the curved helical fractal and linear helical fractal on the other hand, this is not true.}
\label{fig10}
\end{figure}

For the curved helical fractal and linear helical fractal on the other hand, this may not be true since they are not self-similar. We assumed that $D_{\textrm{upper}}$ would be constrained by the crowding at the inner regions of the circle and one winding of the helix respectively, but will the crowding at finer scales self-intersect for such $D_{\textrm{upper}}$? The answer turns out to be no, and here is why.

First consider one winding of the toroidal curve for the curved helical fractal (the next level after the circle) and the crowding at its inner region. The corresponding value of $D_{\textrm{upper}}$ would be the result of the corresponding dotted circles $A$ to $F$ to be just touching. The arrangements of these dotted circles are similar to those in the helix (displaced along the direction of the circle, and then tilted). Since the circle has length $2\pi$, one complete winding of the toroidal curve would experience a displacement of $2\pi/\omega$. Therefore, two such adjacent dotted circles have a displacement of $2\pi/\omega^2$. The upper bound to $D$ at this scale would be when these displaced and tilted dotted circles are touching, so the Pythagorean theorem tells us that
\begin{eqnarray}\label{generic}
\left(2r''_\infty\right)^2=s''^2+\left(\frac{2\pi}{\omega^2}\right)^2,
\end{eqnarray}
where  $r''_\infty=\displaystyle\sum_{i=3}^\infty{R^i}=\frac{R^3}{1-R}$ and $s''=2\mu R(1-R)$ for this scale. This is exactly the same as Eq. (\ref{preDupperhelix}) that arose when we calculated $D_{\textrm{upper}}$ for the helix. We have already explained that this will give a value of $D_{\textrm{upper}}$ that is always larger than that in Eq. (\ref{Dupperrr}) because of the additional displacement that provides more space. The same kind of additional displacement exists at even finer scales of the curved helical fractal. Conclusively, the least upper bound to the Hausdorff dimension for the curved helical fractal is indeed given by Eq. (\ref{Dupperrr}).

To reach a similar conclusion for the helical fractal, note that the equation corresponding to Eq. (\ref{preDupperhelix}) for finer scales (i.e. crowding at the inner region of one winding of the $n$-th level) is
\begin{eqnarray}
\left(2r^{(n)}_\infty\right)^2=s_n^2+a_n^2,
\end{eqnarray}
where $\displaystyle r^{(n)}_\infty=\sum_{i=n+1}^{\infty}{R^i}=\frac{R^{n+1}}{1-R}$, $s_n=2\mu R^{n-1}(1-R)$, $\displaystyle a_n=\frac{1}{\omega^2}\left(\frac{L_{n-1}}{\omega^{n-2}}\right)$ is the displacement between two adjacent dotted circles along $\vec{H}_{n-1}$. (Note that $L_{n-1}/\omega^{n-2}$ represents the length of one winding of the $(n-1)$-th level.) Some simplifications would lead to the equation of the following form:
\begin{eqnarray}
\left(2r'_\infty\right)^2=s'^2+\left(\frac{2\pi b}{\omega^2}\right)^2,
\end{eqnarray}
where $b\geq1$, and $b=1$ only in the case of $n=2$ which is at the scale of the helix (as derived previously). This implies that the displacements occurring at finer scales are larger than that in Eq. (\ref{preDupperhelix}), yielding more space and would lead to a bigger value for $D_{\textrm{upper}}$. Hence, the least upper bound to the Hausdorff dimension for the linear helical fractal is indeed the one calculated from Eqs. (\ref{Dupperhelix}) and (\ref{D}), shown in Fig. 8. It is worth pointing out that the non-exact self-similarity of the linear helical fractal is what leads to this subtlety, where the no-self-intersection constraint is the tightest at the scale of the helix.

\bibliographystyle{spphys}       
\bibliography{Citation}

\begin{thebibliography}{10}
\providecommand{\url}[1]{{#1}}
\providecommand{\urlprefix}{URL }
\expandafter\ifx\csname urlstyle\endcsname\relax
  \providecommand{\doi}[1]{DOI \discretionary{}{}{}#1}\else
  \providecommand{\doi}{DOI \discretionary{}{}{}\begingroup
  \urlstyle{rm}\Url}\fi

\bibitem{fle1}
N.~Fletcher, T.~Tarnopolskaya, F.~de~Hoog, Proceedings of the Royal Society of
  London. Series A: Mathematical, Physical and Engineering Sciences
  \textbf{457}(2005), 33 (2001).
\newblock \doi{10.1098/rspa.2000.0654}.
\newblock
  \urlprefix\url{http://rspa.royalsocietypublishing.org/content/457/2005/33.abstract}

\bibitem{fle2}
N.H. Fletcher, American Journal of Physics \textbf{72}(5), 701 (2004).
\newblock \doi{10.1119/1.1652038}.
\newblock \urlprefix\url{http://link.aip.org/link/?AJP/72/701/1}

\bibitem{tolsua1}
C.D. Toledo-Su\'arez, Chaos, Solitons \& Fractals \textbf{39}(1), 342  (2009).
\newblock \doi{10.1016/j.chaos.2007.01.095}.
\newblock
  \urlprefix\url{http://www.sciencedirect.com/science/article/pii/S0960077907002068}

\bibitem{text1}
H.G. Schuster, W.~Just, \emph{Deterministic Chaos: An Intro} (Wiley and Sons,
  2005)

\bibitem{bio1}
A.~Annunziato, Nature Education \textbf{1(1)}, 26 (2008)

\bibitem{bio2}
B.A. Pierce, \emph{Genetics: A Conceptual Approach} (W. H. Freeman, 2007)

\bibitem{bio3}
D.E. Olins, A.L. Olins, Nature Reviews Molecular Cell Biology \textbf{4}, 809
  (2003)

\bibitem{bio4}
R.M. Youngson, \emph{Collins dictionary of human biology} (Collins, Glasgow,
  Scotland, 2006)

\bibitem{DNA1}
A.~Provata, Y.~Almirantis, Fractals \textbf{08}(01), 15 (2000).
\newblock \doi{10.1142/S0218348X00000044}.
\newblock
  \urlprefix\url{http://www.worldscientific.com/doi/abs/10.1142/S0218348X00000044}

\bibitem{DNA2}
C.K. Peng, S.V. Buldyrev, A.L. Goldberger, S.~Havlin, F.~Sciortino, M.~Simons,
  S.H. E., Nature \textbf{356}({6365}), 168 (1992).
\newblock \doi{{10.1038/356168a0}}

\bibitem{Prot1}
T.G. Dewey, Fractals \textbf{01}(02), 179 (1993).
\newblock \doi{10.1142/S0218348X93000198}.
\newblock
  \urlprefix\url{http://www.worldscientific.com/doi/abs/10.1142/S0218348X93000198}

\bibitem{Prot2}
E.~Tejera, A.~Machado, I.~Rebelo, J.~Nieto-Villar, Physica A: Statistical
  Mechanics and its Applications \textbf{388}(21), 4600  (2009).
\newblock \doi{http://dx.doi.org/10.1016/j.physa.2009.07.015}.
\newblock
  \urlprefix\url{http://www.sciencedirect.com/science/article/pii/S0378437109005792}

\bibitem{Prot3}
X.~Peng, W.~Qi, M.~Wang, R.~Su, Z.~He, Communications in Nonlinear Science and
  Numerical Simulation \textbf{18}(12), 3373  (2013).
\newblock \doi{http://dx.doi.org/10.1016/j.cnsns.2013.05.005}.
\newblock
  \urlprefix\url{http://www.sciencedirect.com/science/article/pii/S1007570413002074}

\bibitem{Prot4}
D.~Craciun, A.~Isvoran, R.D. Reisz, N.M. Avram, Fractals \textbf{18}(02), 207
  (2010).
\newblock \doi{10.1142/S0218348X10004798}.
\newblock
  \urlprefix\url{http://www.worldscientific.com/doi/abs/10.1142/S0218348X10004798}

\bibitem{Prot5}
A.A. Kaczor, R.~Guixa -Gonzalez, P.~Carrio, C.~Obiol-Pardo, M.~Pastor,
  J.~Selent, Journal of Molecular Modeling \textbf{18}(9), 4465 (2012).
\newblock \doi{10.1007/s00894-012-1431-2}.
\newblock \urlprefix\url{http://dx.doi.org/10.1007/s00894-012-1431-2}

\bibitem{text2}
S.H. Strogatz, \emph{Nonlinear Dynamics and Chaos: With applications to
  physics, biology, chemisty, and engineering} (Westview Press, 2001)

\bibitem{Pre}
A.N. Pressley, \emph{Elementary differential geometry} (Springer, New York,
  2008)

\bibitem{topo1}
M.~Nakahara, \emph{Geometry, topology and physics} (Taylor and Francis, 2003)

\bibitem{topo2}
A.~Hatcher, \emph{Algebraic topology} (Cambridge University Press, 2001)

\bibitem{knot1}
W.~Lickorish, \emph{An introduction to knot theory} (Springer, 1997)

\bibitem{knot2}
K.~Murasugi, \emph{Knot theory and its applications} (Birkh\"{a}user, 2007)

\bibitem{knot3}
S.~Chmutov, S.~Duzhin, J.~Mostovoy, \emph{Introduction to Vassiliev knot
  invariants} (Cambridge University Press, 2012)

\bibitem{Vee2012}
V.-L. Saw, L.Y. Chew, Gen. Relativ. and Gravit. \textbf{44}, 2989 (2012).
\newblock \doi{10.1007/s10714-012-1435-3}.
\newblock \urlprefix\url{http://dx.doi.org/10.1007/s10714-012-1435-3}

\bibitem{Vee2013}
V.-L. Saw, L.Y. Chew, Gen. Relativ. and Gravit. \textbf{46}, 1655 (2014).
\newblock \doi{10.1007/s10714-013-1655-1}.
\newblock \urlprefix\url{http://dx.doi.org/10.1007/s10714-013-1655-1}

\bibitem{Vee2014}
V.-L. Saw, "A rotating universe outside a Schwarzschild black hole where
  spacetime itself non-uniformly rotates", arXiv:1403.0337 [gr-qc]  (2014).
\newblock \urlprefix\url{http://arxiv.org/abs/1403.0337}

\end{thebibliography}

\end{document}